\documentclass[9pt,twocolumn,twoside]{osajnl}

\journal{ol} 

\setboolean{shortarticle}{true}

\title{Optical Two-dimensional Coherent Spectroscopy of Cold Atoms}

\author[1]{Danfu Liang}
\author[1]{Lexter Savio Rodriguez}
\author[1]{Haitao Zhou}
\author[1]{Yifu Zhu}
\author[1,*]{Hebin Li}

\affil[1]{Department of physics, Florida International University, Miami, FL. 33199, USA}

\affil[*]{Corresponding author: hebin.li@fiu.edu}

\begin{abstract}

We report an experimental demonstration of optical 2DCS in cold atoms. The experiment integrates a collinear 2DCS setup with a magneto-optical trap (MOT), in which cold rubidium (Rb) atoms are prepared at a temperature of about 200 $\mu$K and a number density of $10^{10}$ cm$^{-3}$. With a sequence of femtosecond laser pulses, we first obtained one-dimensional second- and fourth-order nonlinear signals and then acquired both one-quantum and zero-quantum 2D spectra of cold Rb atoms. The capability of performing optical 2DCS in cold atoms is an important step toward optical 2DCS study of many-body physics in cold atoms and ultimately in atom arrays and trapped ions. Optical 2DCS in cold atoms/molecules can also be a new avenue to probe chemical reaction dynamics in cold molecules.

\end{abstract}

\begin{document}

\maketitle
Ultrafast femtosecond lasers usually are not used for studies in cold atoms since the two systems are apparently incompatible in their characteristic time scales. A few experiments have utilized ultrafast lasers to demonstrate ultrafast manipulation of quantum states and entanglement in trapped ions \cite{PhysRevLett.110.203001,PhysRevLett.119.230501} and neutral atoms \cite{PhysRevA.97.052322,https://doi.org/10.48550/arxiv.2111.12314}, enabling the possibility of atom-based quantum gate operation at picosecond or shorter time scales. The high intensity of femtosecond pulses also provides an opportunity to perform ultrafast nonlinear spectroscopy in cold atoms which has been largely overlooked. Here we report the implementation of an advanced ultrafast spectroscopic technique, namely optical two-dimensional coherent spectroscopy (2DCS), in a cold atom cloud prepared in a magneto-optical trap (MOT).

Originated in nuclear magnetic resonance \cite{Ernstbook}, the concept of multi-dimensional Fourier-transform spectroscopy has been implemented in the optical region using femtosecond lasers and developed into a powerful tool to study energy level structures, couplings, and dynamics in a variety of complex systems such as proteins \cite{Hamm1999a}, photosynthetic systems \cite{doi:10.1126/sciadv.aaz4888,doi:10.1073/pnas.1702261114}, semiconductor quantum wells \cite{Cundiff2012,Li2006a,Nardin2014,Singh2013,Turner2012}, quantum dots \cite{Moody2013b,Moody2013a,Moody2013,PhysRevB.87.041304}, 2D materials \cite{Moody2015,Titze2018}, perovskites \cite{Thouin2018,Titze2019}, atomic vapors \cite{Tian2003,Dai2010,Dai2012a,Dai2012,Li2013a,Gao:16,PhysRevLett.120.233401,Yu2018,Yu2019,Binz2020,Liang2021,Yu2022,Liang2022,PhysRevA.105.052810}, and weakly-bound molecules on helium nanodroplets in a molecular beam \cite{Bruder2018}. Particularly, double-quantum 2DCS provids an extremely sensitive background-free detection of dipole-dipole interactions in both potassium (K) \cite{Dai2012,Yu2019} and rubidium (Rb) \cite{Gao:16,Yu2019} atomic vapors. The technique can be extended to multi-quantum 2DCS which probe multi-atom correlated coherent states (Dicke states) with a scalable and deterministic number of atoms, up to eight atoms \cite{Yu2019,Liang2021} in a K atomic vapor. The collective states of higher excited states ($D$ state), in addition to $P$ state, has also been observed \cite{Liang2022} by using double-quantum 2DCS. These studies have demonstrated optical 2DCS as a powerful technique to study many-body interactions and correlations in atomic systems.

However, atomic vapors at room or higher temperatures pose extra challenges in the quantitative and deterministic study of many-body interactions and correlations due to the presence of thermal motion. For example, although the number density and the mean interatomic separation are known at a given cell temperature, the exact separation and number of atoms have a statistical distribution but are not deterministic for each shot of the experiment. The thermal motion also speeds up the decoherence of many-body states, making it difficult to form and observe coherent states with a large number of atoms. In contrast, cold atoms provide a well-controlled and isolated environment with virtually no thermal motion. Especially, recent advances in trapping neutral atoms in an array of optical tweezers \cite{Kaufman2014,Lester2015,Kim2016,Barredo2016,Endres2016,Kaufman2021} open the possibility of studying many-body physics in an atom array with a deterministic number of atoms and spatial distribution. Demonstrating optical 2DCS in a cold atom cloud is an essential first step toward this goal. Recent progress in optical 2DCS made it feasible for 2DCS studies in cold atoms. The sensitivity has been improved such that the dipole-dipole interaction can be detected at a density ($10^8$ cm$^{-3}$) lower than a typical cold-atom density \cite{Yu2019}. Optical frequency-comb-based 2DCS has achieved a frequency resolution that is sufficient to resolve all hyperfine levels in Rb and K atoms \cite{Lomsadze2017b,Lomsadze2018}.

\begin{figure*}
\centering
\includegraphics[width=0.9\textwidth]{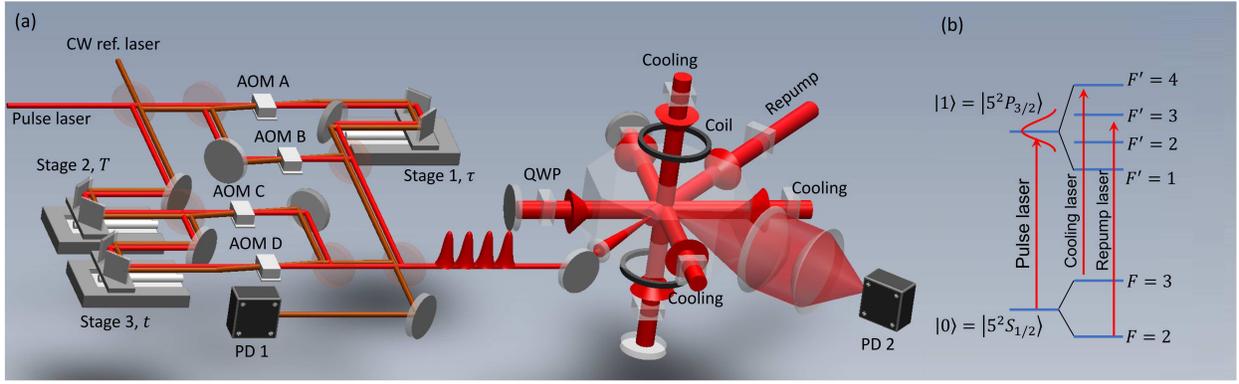}
\caption{(a) Schematic of the experimental setup integrating a collinear 2DCS setup with an Rb MOT. AOM: acousto-optic modulator. QWP: quarter wave plate. PD: photodetector. (b) Relevant energy levels of Rb atoms at the $D_2$ transition and the frequencies of the lasers used in the experiment. The hyperfine splittings are not to scale.
}\label{fig:1}
\end{figure*}

In this letter, we demonstrate the implementation of optical 2DCS in cold Rb atoms by combining a collinear 2DCS setup with an MOT. Rubidium atoms are laser cooled and trapped in the MOT at a temperature of $\sim 200$ $\mu$K. The cold atom cloud includes approximately one million atoms at a number density of $10^{10}$ cm$^{-3}$. With the excitation pulses provided by the collinear 2DCS setup, we first performed one-dimensional (1D) pump-probe and transient four-wave mixing (FWM) spectroscopy. We then demonstrated optical 2DCS and acquired one-quantum and zero-quantum 2D spectra of cold Rb atoms.

The experimental schematic is illustrated in Fig. \ref{fig:1}(a). The system integrates a collinear 2DCS setup (left) with an MOT (right). Four femtosecond pulses derived from the collinear 2DCS setup are incident on a cold atom cloud prepared in the MOT to perform 2DCS measurements. The MOT is constructed in an octagonal glass cell in which the pressure is maintained $<10^{-8}$ torr. An Rb dispenser provides both Rb isotopes at the natural abundance. Our experiment cools and traps $^{85}$Rb atoms. The relevant energy levels and the laser cooling scheme are shown in Fig.\ref{fig:1}(b). The cooling laser with an output power of 200 mW is tuned to the transition $|5^2S_{1/2}, F=3\rangle\rightarrow|5^2P_{3/2}, F^\prime=4\rangle$. The repump laser with an output power of 30 mW is tuned to the transition $|5^2S_{1/2}, F=2\rangle\rightarrow|5^2P_{3/2}, F^\prime=3\rangle$. The cooling laser is divided into three beams and circularly polarized. The three beams converge at the center of the glass cell in three orthogonal directions, two horizontal and one vertical. Each beam is reflected back by a mirror so there are two counter-propagating beams in each dimension for Doppler cooling. The repump laser is also incident on the crossing of cooling laser beams to pump atoms out of $|5^2S_{1/2}, F=2\rangle$ state to maintain the cooling cycle. The laser-cooled atoms are trapped in a magnetic trap generated by a pair of anti-Helmholtz coils. In our experiment, the Rb atoms are cooled to a temperature of $\sim$ 200 $\mu$K and the cold atom gas has a number density of $10^{10}$ cm$^{-3}$.      

With cold atoms prepared in the MOT, optical 2DCS experiment is performed by using a femtosecond laser tuned to excite the $D_2$ transition from $|0\rangle=|5^2S_{1/2}\rangle$ to $|1\rangle=|5^2P_{3/2}\rangle$.
The femtosecond pulse is about $200$ fs in duration at a repetition rate of $78$ MHz. The laser spectrum has a central wavelength of $780$ nm and a bandwidth of $2.6$ nm (the standard deviation). The collinear 2DCS setup consists of three nested Mach-Zehnder interferometers \cite{Nardin2013}, as shown in Fig. \ref{fig:1}(a). The femtosecond pulse is split into four pulses ($A, B, C,$ and $D$) which each go through an acousto-optic modulator (AOM) and are subsequently combined into one beam. The three time delays ($\tau$, $T$ and $t$) between the four pulses are controlled by three translation delay stages. The four pulses are modulated by AOMs at slightly different frequencies of $\Omega_A=80.107$ MHz, $\Omega_B=80.104$ MHz, $\Omega_C=80.0173$ MHz, and $\Omega_D=80$ MHz. At the same time, a continuous-wave (CW) reference laser copropagates with the pulse laser and is modulated by the AOMs in each arm of the interferometer. The combined reference laser beam is detected by a photodetector (PD 1) to provide beat notes of the AOM frequencies. The femtosecond pulses are incident on the cold atom cloud and generate nonlinear fluorescence signals which are detected by another photodetector (PD 2) and a lock-in amplifier. The nonlinear signal induced by a specific pulse sequence can be measured by lock-in detection with a corresponding reference frequency obtained by mixing beat notes of the CW reference laser beams in a digital wave mixer.

\begin{figure}[hbt]
\centering
\includegraphics[width=\columnwidth]{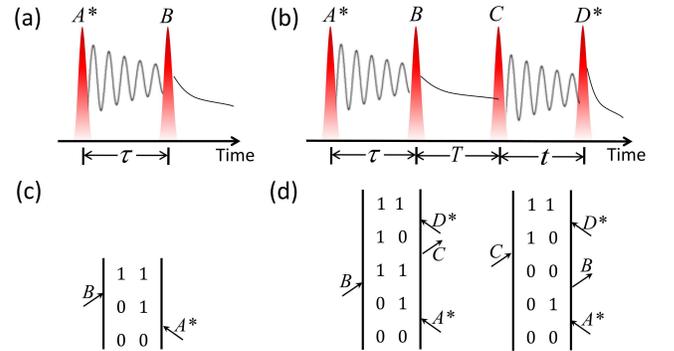}
\caption{Pulse sequences for (a) two-pulse pump-probe spectroscopy and (b) four-pulse excitation for optical 2DCS. Double-side Feynman diagrams describing contributing quantum pathways by (c) two-pulse excitation and (d) four-pulse excitation. }\label{fig:2}
\end{figure} 

Two different sequences of femtosecond pulses are used in the experiment. A two-pulse sequence shown in Fig. \ref{fig:2}(a) is used for pump-probe spectroscopy and a four-pulse sequence shown in Fig. \ref{fig:2}(b) is used for optical 2DCS. The pulses with an asterisk represent that the pulse is considered conjugated, corresponding to a negative AOM modulation frequency. The pump-probe spectroscopy can be performed with any pair of pulses while the other two pulses are blocked. This ensures a good pair-wise pulse overlap in both spatially and temporally before starting the four-pulse experiment. Using pulses $A^*$ and $B$ as an example, the lock-in amplifier detects the fluorescence signal that is modulated at the reference frequency $\Omega_{R0}=-\Omega_A+\Omega_B$. This signal is resulted from the excitation pathway represented by the double-sided Feynman diagram in Fig. \ref{fig:2}(c). The four-pulse sequence in Fig. \ref{fig:2}(b) performs a fourth-order excitation and the corresponding excitation pathways are represented by double-sided Feynman diagrams in Fig. \ref{fig:2}(d). In these processes, the first pulse $A^*$ creates a first-order coherence between states $|0\rangle$ and $|1\rangle$, which evolves during time delay $\tau$. The second pulse $B$ converts the first-order coherence into a population in either $|0\rangle$ or $|1\rangle$ depending on the relative phase. The third pulse $C$ generates a third-order coherence between states $|1\rangle$ and $|0\rangle$, which evolves during time delay $t$. Finally, the fourth pulse $D^*$ converts the third-order coherence into a population in state $|1\rangle$ or $|0\rangle$. The population in $|0\rangle$ does not emit fluorescence so the corresponding pathways are not shown. The fourth-order population in state $|1\rangle$ decays and emits a fourth-order fluorescence signal, which can be detected by a lock-in amplifier referenced to the frequency $\Omega_{R1}= -\Omega_A+\Omega_B + \Omega_C -\Omega_D$. This fourth-order nonlinear signal can be measured as a function of $\tau$, $T$, or/and $t$. When the signal is measured with two varying time delays, a 2D spectrum can be generated by Fourier transforming the time-domain signal with respect to two time delays.

\begin{figure}
\centering
\includegraphics[width=0.9\columnwidth]{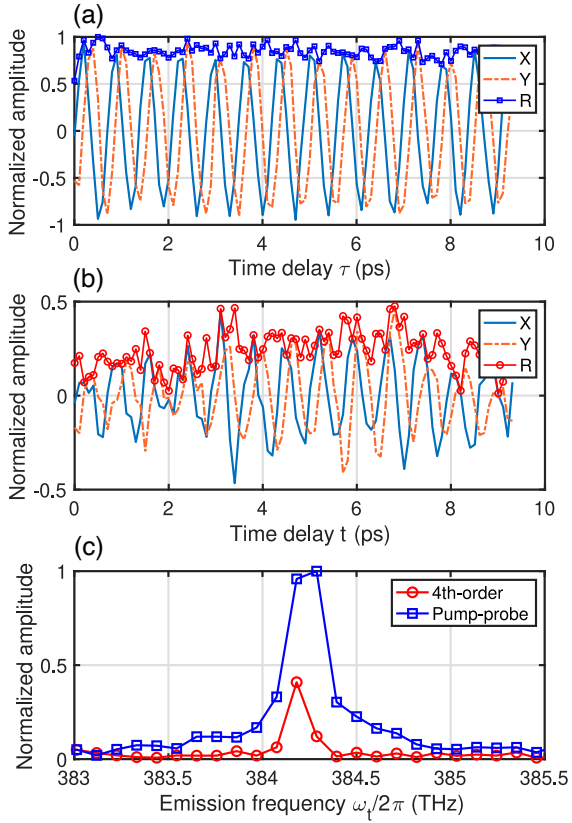}
\caption{One-dimensional time-domain spectra for (a) two-pulse pump-probe experiment and (b) the fourth-order signal due to four-pulse excitation. Real part $X$, imaginary part $Y$, and amplitude $R=\sqrt{X^2+Y^2}$ are plotted. (c) Frequency-domain spectra obtained by Fourier transforming the time-domain pump-probe (blue) and fourth-order (red) signals. }\label{fig:3}
\end{figure}

We first performed one-dimensional (1D) scans with two and four pulses. In the two-pulse experiment, the signal is measured by lock-in detection at the reference frequency $\Omega_{R0}=-\Omega_A+\Omega_B$ as the time delay between $A^*$ and $B$ is scanned. In the four-pulse experiment, the signal measured at the reference frequency $\Omega_{R1}= -\Omega_A+\Omega_B + \Omega_C -\Omega_D$ as the time delay $t$ is scanned while the other two delays are fixed at $\tau=500$ fs and $T=500$ fs. The obtained 1D time-domain spectra are shown in Fig. \ref{fig:3}(a) for two-pulse excitation and Fig. \ref{fig:3} for four-pulse excitation. The $X$ and $Y$ components are real and imaginary parts, respectively, of the lock-in amplifier output and the magnitude of the signal is $R=\sqrt{X^2+Y^2}$. Both 1D spectra reveal the dynamics of coherence $\rho_{01}$. The amplitude of oscillations shows virtually no decay within the time window of 10 ps since the coherence time is expected to be much longer at the ns scale. The signal does not oscillate at the optical carrier frequency, but at a reduced frequency $\nu^*=|\nu_{sig}-\nu_{ref}|$ which is the difference between the signal frequency $\nu_{sig}$ and the CW reference laser frequency $\nu_{ref}$. Fourier transforming these time domain signals gives the frequency-domain spectra in Fig. \ref{fig:3}(c), where the blue curve is the pump-probe spectrum and the red curve is the fourth-order nonlinear spectrum. The frequency axis shows the optical frequency calculated from the reduced frequency as $\nu_{sig}=\nu^*+\nu_{ref}$. Both frequency-domain spectra have a resonance at $384.2$ THz which matches the $D_2$ transition frequency. The linewidth of the spectra is limited by the scanned delay time of 10 ps. The amplitude of the pump-probe spectrum is greater than that of the fourth-order signal since the pump-probe experiment measures a second-order nonlinear response.

\begin{figure}[hbt]
\centering
\includegraphics[width=\columnwidth]{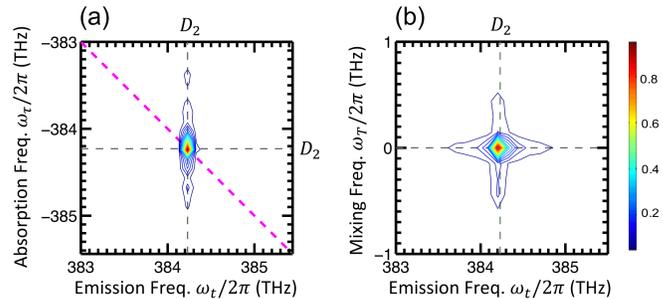}
\caption{(a) One-quantum 2D spectrum and (b) zero-quantum 2D spectrum of cold Rb atoms. The amplitude of the spectra is plotted with the maximum normalized to one.
}\label{fig:4}
\end{figure}

Optical 2DCS measurements can be performed by scanning two time delays while the third time delay is fixed. A one-quantum 2D spectrum, as shown in \ref{fig:3}(a), was obtained by scanning time delays $\tau$ and $t$ with $T=500$ fs. Fourier transform of two time axes generates two frequency axes: absorption frequency $\omega_\tau$ and emission frequency $\omega_t$ corresponding to $\tau$ and $t$, respectively. The resulting 2D peak is located on the diagonal line at the $D_2$ transition frequency of $384.2$ THz. A zero-quantum 2D spectrum, as shown in \ref{fig:3}(b), was obtained by scanning time delays $T$ and $t$ with $\tau=500$ fs. The mixing frequency $\omega_T$ and the emission frequency $\omega_t$ correspond to time delays $T$ and $t$, respectively. The zero-quantum 2D peak has an emission frequency of $384.2$ THz while the mixing frequency is zero since the population only has an exponential decay with a lifetime of $26.2$ ns during time delay $T$. Both 2D spectra are plotted with 20 contours with no visible background noise, demonstrating excellent signal-to-noise ratio. The spectral resolution is $0.1$ THz limited by the scanned time delay of $10$ ps so the hyperfine structures are not resolved. The spectral resolution can be improved by scanning longer delays. However, it is not practical to have a long enough delay to achieve a sufficient frequency resolution ($29$ MHz) to fully resolve hyperfine structures. Optical 2DCS with such frequency resolution is possible by utilizing femtosecond frequency combs \cite{Lomsadze2017b,Lomsadze2018}. The obtained 2D spectra of cold atoms are consistent with the previous experiments \cite{Tian2003,PhysRevA.105.052810} performed in hot Rb atomic vapor, demonstrating the capability of perform optical 2DCS in cold atoms.

In summary, we have experimentally demonstrated the feasibility of optical 2DCS in cold atom clouds prepared in an MOT setup. We first probed 1D second-order and fourth-order nonlinear signals to confirm that high-order nonlinear signals can be generated in cold atoms by femtosecond laser pulses while maintaining a stable cold atom cloud. We then extended the experiment to 2DCS by scanning two time delays. Both one-quantum and zero-quantum 2D spectra were acquired and the results are consistent with previous experiments in atomic vapors. This measurement is extremely robust in probing a specific high-order nonlinear process in cold atoms. This demonstration is an important first step towards the goal of using optical 2DCS to study many-body physics in cold atoms and ultimately in atom arrays trapped by optical tweezers. Similarly, the technique also makes it possible to realize optical 2DCS experiments that have been proposed to study quantum entanglements in trapped ions \cite{37,38}. Ultrafast spectroscopy has been traditionally used to study chemical reactions, optical 2DCS in cold atoms/molecules will open a new avenue to study reaction dynamics in ultracold molecular chemical reactions \cite{Hu2019}.

\medskip

\noindent\textbf{Funding.} National Science Foundation (PHY-2216824). D.L. acknowledges the support by a Dissertation Year Fellowship from FIU. 

\medskip

\noindent\textbf{Disclosures.} The authors declare no conflicts of interest.



\bibliography{References}

\begin{thebibliography}{10}
\newcommand{\enquote}[1]{``#1''}

\bibitem{PhysRevLett.110.203001}
J.~Mizrahi, C.~Senko, B.~Neyenhuis, K.~G. Johnson, W.~C. Campbell, C.~W.~S.
  Conover, and C.~Monroe, \enquote{Ultrafast spin-motion entanglement and
  interferometry with a single atom,} {\protect\JournalTitle{Phys. Rev. Lett.}}
  \textbf{110}, 203001 (2013).

\bibitem{PhysRevLett.119.230501}
J.~D. Wong-Campos, S.~A. Moses, K.~G. Johnson, and C.~Monroe,
  \enquote{Demonstration of two-atom entanglement with ultrafast optical
  pulses,} {\protect\JournalTitle{Phys. Rev. Lett.}} \textbf{119}, 230501
  (2017).

\bibitem{PhysRevA.97.052322}
Y.~Song, H.-g. Lee, H.~Kim, H.~Jo, and J.~Ahn, \enquote{Subpicosecond $x$
  rotations of atomic clock states,} {\protect\JournalTitle{Phys. Rev. A}}
  \textbf{97}, 052322 (2018).

\bibitem{https://doi.org/10.48550/arxiv.2111.12314}
Y.~Chew, T.~Tomita, T.~P. Mahesh, S.~Sugawa, S.~de~Léséleuc, and K.~Ohmori,
  \enquote{Ultrafast energy exchange between two single rydberg atoms on the
  nanosecond timescale,} {\protect\JournalTitle{arXiv:2111.12314}}  (2021).

\bibitem{Ernstbook}
R.~Ernst, G.~Bodenhausen, and A.~Wokaun, \emph{{Principles of Nuclear Magnetic
  Resonance in One and Two Dimensions}} (Oxford Science Publications, Oxford,
  U.K., 1987).

\bibitem{Hamm1999a}
P.~Hamm, M.~Lim, W.~F. DeGrado, and R.~M. Hochstrasser, \enquote{{The
  two-dimensional IR nonlinear spectroscopy of a cyclic penta-peptide in
  relation to its three-dimensional structure},} {\protect\JournalTitle{Proc.
  Natl. Acad. Sci. U.S.A}} \textbf{96}, 2036--2041 (1999).

\bibitem{doi:10.1126/sciadv.aaz4888}
J.~Cao, R.~J. Cogdell, D.~F. Coker, H.-G. Duan, J.~Hauer, U.~Kleinekathöfer,
  T.~L.~C. Jansen, T.~Mančal, R.~J.~D. Miller, J.~P. Ogilvie, V.~I.
  Prokhorenko, T.~Renger, H.-S. Tan, R.~Tempelaar, M.~Thorwart, E.~Thyrhaug,
  S.~Westenhoff, and D.~Zigmantas, \enquote{Quantum biology revisited,}
  {\protect\JournalTitle{Science Advances}} \textbf{6}, eaaz4888 (2020).

\bibitem{doi:10.1073/pnas.1702261114}
H.-G. Duan, V.~I. Prokhorenko, R.~J. Cogdell, K.~Ashraf, A.~L. Stevens,
  M.~Thorwart, and R.~J.~D. Miller, \enquote{Nature does not rely on long-lived
  electronic quantum coherence for photosynthetic energy transfer,}
  {\protect\JournalTitle{Proceedings of the National Academy of Sciences}}
  \textbf{114}, 8493--8498 (2017).

\bibitem{Cundiff2012}
S.~T. Cundiff, A.~D. Bristow, M.~Siemens, H.~Li, G.~Moody, D.~Karaiskaj,
  X.~Dai, and T.~Zhang, \enquote{{Optical 2-d fourier transform spectroscopy of
  excitons in semiconductor nanostructures},} {\protect\JournalTitle{IEEE J.
  Sel. Top. Quantum Electron.}} \textbf{18}, 318--328 (2012).

\bibitem{Li2006a}
X.~Li, T.~Zhang, C.~Borca, and S.~Cundiff, \enquote{{Many-Body Interactions in
  Semiconductors Probed by Optical Two-Dimensional Fourier Transform
  Spectroscopy},} {\protect\JournalTitle{Phys. Rev. Lett.}} \textbf{96}, 057406
  (2006).

\bibitem{Nardin2014}
G.~Nardin, G.~Moody, R.~Singh, T.~M. Autry, H.~Li, F.~Morier-Genoud, and S.~T.
  Cundiff, \enquote{{Coherent Excitonic Coupling in an Asymmetric Double InGaAs
  Quantum Well Arises from Many-Body Effects},} {\protect\JournalTitle{Phys.
  Rev. Lett.}} \textbf{112}, 046402 (2014).

\bibitem{Singh2013}
R.~Singh, T.~M. Autry, G.~Nardin, G.~Moody, H.~Li, K.~Pierz, M.~Bieler, and
  S.~T. Cundiff, \enquote{{Anisotropic homogeneous linewidth of the heavy-hole
  exciton in (110)-oriented GaAs quantum wells},} {\protect\JournalTitle{Phys.
  Rev. B}} \textbf{88}, 45304 (2013).

\bibitem{Turner2012}
D.~Turner, P.~Wen, D.~Arias, K.~Nelson, H.~Li, G.~Moody, M.~Siemens, and
  S.~Cundiff, \enquote{{Persistent exciton-type many-body interactions in GaAs
  quantum wells measured using two-dimensional optical spectroscopy},}
  {\protect\JournalTitle{Phys. Rev. B}} \textbf{85}, 201303 (2012).

\bibitem{Moody2013b}
G.~Moody, R.~Singh, H.~Li, I.~A. Akimov, M.~Bayer, D.~Reuter, A.~D. Wieck,
  A.~S. Bracker, D.~Gammon, and S.~T. Cundiff, \enquote{{Biexcitons in
  semiconductor quantum dot ensembles},} {\protect\JournalTitle{Phys. Status
  Solidi (b)}} \textbf{250}, 1753--1759 (2013).

\bibitem{Moody2013a}
G.~Moody, R.~Singh, H.~Li, I.~Akimov, M.~Bayer, D.~Reuter, A.~Wieck, and
  S.~Cundiff, \enquote{{Correlation and dephasing effects on the non-radiative
  coherence between bright excitons in an InAs QD ensemble measured with 2D
  spectroscopy},} {\protect\JournalTitle{Solid State Commun}} \textbf{163},
  65--69 (2013).

\bibitem{Moody2013}
G.~Moody, R.~Singh, H.~Li, I.~A. Akimov, M.~Bayer, D.~Reuter, A.~D. Wieck, and
  S.~T. Cundiff, \enquote{{Fifth-order nonlinear optical response of excitonic
  states in an InAs quantum dot ensemble measured with two-dimensional
  spectroscopy},} {\protect\JournalTitle{Phys. Rev. B}} \textbf{87}, 045313
  (2013).

\bibitem{PhysRevB.87.041304}
G.~Moody, R.~Singh, H.~Li, I.~A. Akimov, M.~Bayer, D.~Reuter, A.~D. Wieck,
  A.~S. Bracker, D.~Gammon, and S.~T. Cundiff, \enquote{{Influence of
  confinement on biexciton binding in semiconductor quantum dot ensembles
  measured with two-dimensional spectroscopy},} {\protect\JournalTitle{Phys.
  Rev. B}} \textbf{87}, 41304 (2013).

\bibitem{Moody2015}
G.~Moody, C.~{Kavir Dass}, K.~Hao, C.-H. Chen, L.-J. Li, A.~Singh, K.~Tran,
  G.~Clark, X.~Xu, G.~Bergh{\"{a}}user, E.~Malic, A.~Knorr, and X.~Li,
  \enquote{{Intrinsic homogeneous linewidth and broadening mechanisms of
  excitons in monolayer transition metal dichalcogenides},}
  {\protect\JournalTitle{Nat. Commun.}} \textbf{6}, 8315 (2015).

\bibitem{Titze2018}
M.~Titze, B.~Li, X.~Zhang, P.~M. Ajayan, and H.~Li, \enquote{{Intrinsic
  coherence time of trions in monolayer MoSe2 measured via two-dimensional
  coherent spectroscopy},} {\protect\JournalTitle{Phys. Rev. Materials}}
  \textbf{2}, 054001 (2018).

\bibitem{Thouin2018}
F.~Thouin, S.~Neutzner, D.~Cortecchia, V.~A. Dragomir, C.~Soci, T.~Salim, Y.~M.
  Lam, R.~Leonelli, A.~Petrozza, A.~R.~S. Kandada, and C.~Silva,
  \enquote{Stable biexcitons in two-dimensional metal-halide perovskites with
  strong dynamic lattice disorder,} {\protect\JournalTitle{Phys. Rev.
  Materials}} \textbf{2} (2018).

\bibitem{Titze2019}
M.~Titze, C.~Fei, M.~Munoz, X.~Wang, H.~Wang, and H.~Li, \enquote{Ultrafast
  carrier dynamics of dual emissions from the orthorhombic phase in
  methylammonium lead iodide perovskites revealed by two-dimensional coherent
  spectroscopy,} {\protect\JournalTitle{J. Phys. Chem. Lett.}} \textbf{10},
  4625--4631 (2019).

\bibitem{Tian2003}
P.~F. Tian, D.~Keusters, Y.~Suzaki, and W.~S. Warren, \enquote{{Femtosecond
  phase-coherent two-dimensional spectroscopy.}}
  {\protect\JournalTitle{Science}} \textbf{300}, 1553--1555 (2003).

\bibitem{Dai2010}
X.~Dai, A.~D. Bristow, D.~Karaiskaj, and S.~T. Cundiff,
  \enquote{Two-dimensional fourier-transform spectroscopy of potassium vapor,}
  {\protect\JournalTitle{Phys. Rev. A}} \textbf{82}, 052503 (2010).

\bibitem{Dai2012a}
X.~Dai, M.~Richter, H.~Li, A.~D. Bristow, C.~Falvo, S.~Mukamel, and S.~T.
  Cundiff, \enquote{{Two-Dimensional Double-Quantum Spectra Reveal Collective
  Resonances in an Atomic Vapor},} {\protect\JournalTitle{Phys. Rev. Lett.}}
  \textbf{108}, 193201 (2012).

\bibitem{Dai2012}
X.~Dai, M.~Richter, H.~Li, A.~D. Bristow, C.~Falvo, S.~Mukamel, and S.~T.
  Cundiff, \enquote{{Two-Dimensional Double-Quantum Spectra Reveal Collective
  Resonances in an Atomic Vapor},} {\protect\JournalTitle{Phys. Rev. Lett.}}
  \textbf{108}, 193201 (2012).

\bibitem{Li2013a}
H.~Li, A.~D. Bristow, M.~E. Siemens, G.~Moody, and S.~T. Cundiff,
  \enquote{{Unraveling quantum pathways using optical 3D Fourier-transform
  spectroscopy.}} {\protect\JournalTitle{Nat. Commun.}} \textbf{4}, 1390
  (2013).

\bibitem{Gao:16}
F.~Gao, S.~T. Cundiff, and H.~Li, \enquote{{Probing dipole--dipole interaction
  in a rubidium gas via double-quantum 2D spectroscopy},}
  {\protect\JournalTitle{Opt. Lett.}} \textbf{41}, 2954--2957 (2016).

\bibitem{PhysRevLett.120.233401}
B.~Lomsadze and S.~T. Cundiff, \enquote{Frequency-comb based double-quantum
  two-dimensional spectrum identifies collective hyperfine resonances in atomic
  vapor induced by dipole-dipole interactions,} {\protect\JournalTitle{Phys.
  Rev. Lett.}} \textbf{120}, 233401 (2018).

\bibitem{Yu2018}
S.~Yu, M.~Titze, Y.~Zhu, X.~Liu, and H.~Li, \enquote{Long range dipole-dipole
  interaction in low-density atomic vapors probed by double-quantum
  two-dimensional coherent spectroscopy,} {\protect\JournalTitle{Opt. Express}}
  \textbf{27}, 28891 (2019).

\bibitem{Yu2019}
S.~Yu, M.~Titze, Y.~Zhu, X.~Liu, and H.~Li, \enquote{{Observation of scalable
  and deterministic multi-atom Dicke states in an atomic vapor},}
  {\protect\JournalTitle{Opt. Lett.}} \textbf{44}, 2795 (2019).

\bibitem{Binz2020}
M.~Binz, L.~Bruder, L.~Chen, M.~F. Gelin, W.~Domcke, and F.~Stienkemeier,
  \enquote{Effects of high pulse intensity and chirp in two-dimensional
  electronic spectroscopy of an atomic vapor,} {\protect\JournalTitle{Opt.
  Express}} \textbf{28}, 25806--25829 (2020).

\bibitem{Liang2021}
D.~Liang and H.~Li, \enquote{Optical two-dimensional coherent spectroscopy of
  many-body dipole–dipole interactions and correlations in atomic vapors,}
  {\protect\JournalTitle{The Journal of Chemical Physics}} \textbf{154}, 214301
  (2021).

\bibitem{Yu2022}
S.~Yu, Y.~Geng, D.~Liang, H.~Li, and X.~Liu,
  \enquote{Double-quantum--zero-quantum 2d coherent spectroscopy reveals
  quantum coherence between collective states in an atomic vapor,}
  {\protect\JournalTitle{Opt. Lett.}} \textbf{47}, 997--1000 (2022).

\bibitem{Liang2022}
D.~Liang, Y.~Zhu, and H.~Li, \enquote{Collective resonance of $d$ states in
  rubidium atoms probed by optical two-dimensional coherent spectroscopy,}
  {\protect\JournalTitle{Phys. Rev. Lett.}} \textbf{128}, 103601 (2022).

\bibitem{PhysRevA.105.052810}
J.~Yan, S.~Revesz, D.~Liang, and H.~Li, \enquote{Broadband optical
  two-dimensional coherent spectroscopy of a rubidium atomic vapor,}
  {\protect\JournalTitle{Phys. Rev. A}} \textbf{105}, 052810 (2022).

\bibitem{Bruder2018}
L.~Bruder, U.~Bangert, M.~Binz, D.~Uhl, R.~Vexiau, N.~Bouloufa-Maafa,
  O.~Dulieu, and F.~Stienkemeier, \enquote{Coherent multidimensional
  spectroscopy of dilute gas-phase nanosystems,} {\protect\JournalTitle{Nature
  Communications}} \textbf{9}, 4823 (2018).

\bibitem{Kaufman2014}
A.~M. Kaufman, B.~J. Lester, C.~M. Reynolds, M.~L. Wall, M.~Foss-Feig, K.~R.~A.
  Hazzard, A.~M. Rey, and C.~A. Regal, \enquote{{Two-particle quantum
  interference in tunnel-coupled optical tweezers},}
  {\protect\JournalTitle{Science}} \textbf{345}, 306--309 (2014).

\bibitem{Lester2015}
B.~J. Lester, N.~Luick, A.~M. Kaufman, C.~M. Reynolds, and C.~A. Regal,
  \enquote{{Rapid Production of Uniformly Filled Arrays of Neutral Atoms},}
  {\protect\JournalTitle{Physical Review Letters}} \textbf{115}, 073003 (2015).

\bibitem{Kim2016}
H.~Kim, W.~Lee, H.-g. Lee, H.~Jo, Y.~Song, and J.~Ahn, \enquote{{In situ
  single-atom array synthesis using dynamic holographic optical tweezers},}
  {\protect\JournalTitle{Nature Communications}} \textbf{7}, 13317 (2016).

\bibitem{Barredo2016}
D.~Barredo, S.~de~L{\'{e}}s{\'{e}}leuc, V.~Lienhard, T.~Lahaye, and
  A.~Browaeys, \enquote{{An atom-by-atom assembler of defect-free arbitrary
  two-dimensional atomic arrays},} {\protect\JournalTitle{Science}}
  \textbf{354}, 1021--1023 (2016).

\bibitem{Endres2016}
M.~Endres, H.~Bernien, A.~Keesling, H.~Levine, E.~R. Anschuetz, A.~Krajenbrink,
  C.~Senko, V.~Vuletic, M.~Greiner, and M.~D. Lukin, \enquote{{Atom-by-atom
  assembly of defect-free one-dimensional cold atom arrays},}
  {\protect\JournalTitle{Science}} \textbf{354}, 1024--1027 (2016).

\bibitem{Kaufman2021}
A.~M. Kaufman and K.-K. Ni, \enquote{{Quantum science with optical tweezer
  arrays of ultracold atoms and molecules},} {\protect\JournalTitle{Nature
  Physics}} \textbf{17}, 1324--1333 (2021).

\bibitem{Lomsadze2017b}
B.~Lomsadze and S.~T. Cundiff, \enquote{{Frequency combs enable rapid and
  high-resolution multidimensional coherent spectroscopy},}
  {\protect\JournalTitle{Science}} \textbf{357}, 1389--1391 (2017).

\bibitem{Lomsadze2018}
B.~Lomsadze and S.~T. Cundiff, \enquote{{Frequency-Comb Based Double-Quantum
  Two-Dimensional Spectrum Identifies Collective Hyperfine Resonances in Atomic
  Vapor Induced by Dipole-Dipole Interactions},} {\protect\JournalTitle{Phys.
  Rev. Lett.}} \textbf{120}, 233401 (2018).

\bibitem{Nardin2013}
G.~Nardin, T.~M. Autry, K.~L. Silverman, and S.~T. Cundiff,
  \enquote{{Multidimensional coherent photocurrent spectroscopy of a
  semiconductor nanostructure},} {\protect\JournalTitle{Opt. Express}}
  \textbf{21}, 28617 (2013).

\bibitem{37}
F.~Schlawin, M.~Gessner, S.~Mukamel, and A.~Buchleitner, \enquote{Nonlinear
  spectroscopy of trapped ions,} {\protect\JournalTitle{Phys. Rev. A}}
  \textbf{90}, 023603 (2014).

\bibitem{38}
M.~Gessner, F.~Schlawin, H.~Häffner, S.~Mukamel, and A.~Buchleitner,
  \enquote{Nonlinear spectroscopy of controllable many-body quantum systems,}
  {\protect\JournalTitle{New Journal of Physics}} \textbf{16}, 092001 (2014).

\bibitem{Hu2019}
M.-G. Hu, Y.~Liu, D.~D. Grimes, Y.-W. Lin, A.~H. Gheorghe, R.~Vexiau,
  N.~Bouloufa-Maafa, O.~Dulieu, T.~Rosenband, and K.-K. Ni, \enquote{{Direct
  observation of bimolecular reactions of ultracold KRb molecules},}
  {\protect\JournalTitle{Science}} \textbf{366}, 1111--1115 (2019).

\end{thebibliography}


\end{document}